\documentclass[aps,prd,onecolumn,groupedaddress,nofootinbib,amssymb]{revtex4-2}
\usepackage[figuresright]{rotating}
\usepackage{amsmath}
\usepackage{xcolor}
\usepackage{graphicx}
\usepackage{amsfonts}
\usepackage{bm}
\usepackage{cancel}
\usepackage{comment}
\usepackage{float}
\newcommand\be{\begin{equation}}
\newcommand\ee{\end{equation}}

\allowdisplaybreaks[4]
\begin{document}

\title{Cosmological perturbations in the spatially covariant gravity with a dynamical lapse function}

	\author{Xue-Zheng Zhu}%
	\email[Email: ]{zhuxzh5@mail2.sysu.edu.cn}
\affiliation{%
	School of Physics and Astronomy, Sun Yat-sen University, Guangzhou 510275, China}

\author{Yang Yu}%
	\email[Email: ]{yuyang69@mail2.sysu.edu.cn}
\affiliation{%
	School of Physics and Astronomy, Sun Yat-sen University, Guangzhou 510275, China}

\author{Xian Gao}%
\email[Email: ]{gaoxian@mail.sysu.edu.cn}
\affiliation{%
	School of Physics and Astronomy, Sun Yat-sen University, Guangzhou 510275, China}

\tolerance=5000
\begin{abstract}
 We investigate the scalar perturbations in a class of spatially covariant gravity theory with a dynamical lapse function.
	Generally, there are two scalar degrees of freedom due to the presence of the velocity of the lapse function.
	We treat the scalar perturbations as analogues of those in a two-field inflationary mode, in which one is light mode and the other is the heavy mode.
	This is justified by the fact that the scalar mode due to the dynamical lapse function becomes infinitely heavy in the limit when the lapse function reduces to be an auxiliary variable.
	The standard approaches of multiple filed perturbations can be applied to deal with our model.
	By integrating out the heavy mode and derive the effective theory for the single light field, we find the solution to the single mode in the form of plane waves.
	Then we  calculate the corrections to the power spectrum of the light mode from the heavy mode, by making use of the standard perturbative method of field theory.
    At last, when the two fields are not weakly coupled, we find a power law mode for the coupled system in large scales.
\end{abstract}



\maketitle

\section{Introduction}

Inflation is one of the most important ideas in modern cosmology \cite{Sato:2015dga}.
In a type model of inflation, one or several scalar fields derive the inflation and generate the quantum fluctuations in the primordial universe.

Inflationary model building also stimulates the study of modified gravity containing
additional degrees of freedom beyond the general relativity (GR) (see \cite{Clifton:2011jh,Joyce:2014kja,Berti:2015itd} for recent reviews).
The Horndeski theory \cite{Horndeski:1974wa,Deffayet:2011gz,Kobayashi:2011nu} is
the most general scalar-tensor theory with a single scalar field, in which both the Lagrangian and the equations of motion are up to the second order in derivatives. In recent years, scalar-tensor theories that generalize the Horndeski theory have also been widely investigated \cite{Gleyzes:2014dya,Gleyzes:2014qga,Langlois:2015cwa,BenAchour:2016fzp,Crisostomi:2016czh}.
An alternative approach to generalizing the scalar-tensor theory is to employ the spatially covariant gravity \cite{Gao:2014soa,Gao:2014fra}, which is a metric theory that respect only the spatial covariance.
Since the general covariance is broken, it is natural that the lapse function is not an auxiliary variable any more, and also acquires a velocity term $\dot{N}$.
This possibility is considered in \cite{Gao:2018znj,Gao:2019lpz}, which show that there will be an extra scalar mode due to the dynamical lapse function.
Usually one has to impose additional conditions in order to remove such an extra mode.
On the other hand, from the point of view of the spatially covariant gravity, since the equations of motion of all the modes are of the second order, the extra mode itself is well-behaved.
The extra mode loses its dynamics when the lapse function becomes an auxiliary variable, and thus can be treated as a ``heavy field''.
More precisely, we treat the usual curvature perturbation $\zeta$ as the light mode, and the extra mode $A$, which becomes dynamical due to the dynamical lapse function, as the heavy mode.
This is the analogue to the well-studied quasi-single field inflation \cite{Chen:2009zp,Chen:2012ge,Noumi:2012vr} as well as more general multiple inflation models in which heavy modes are present \cite{Cespedes:2012hu,Achucarro:2012yr,Burgess:2012dz,Gao:2012uq,Gao:2013ota,Noumi:2013cfa,Gao:2015aba} (see \cite{Chluba:2015bqa} for a review).

Due to the existence of the coupling terms between the two scalar modes, the perturbations are generally not easy to be solved.
In this paper we apply standard methods in multiple field inflation to solving the coupled system and calculate the corrections to the power spectrum of the light mode due to the presence of the heavy mode.
First we integrate the heavy mode and derive the effective theory for the light mode \cite{Cespedes:2012hu,Achucarro:2012yr,Shiu:2011qw}.
We also calculate the influence of the heavy field on the light field by the standard ``in-in formalism'' of field theory.

When the two fields are not  weakly coupled, it will be  difficult to using perturbative method of field theory. In this case, we try to solve this coupling system in large scales \cite{Tolley:2007nq,Koyama:2007ag}.

The rest part of this paper is organized as follows.
In Sec. \ref{sec:pert} we described our mode and the cosmological perturbations.
In Sec. \ref{sec:int} we derived the effective theory for the light mode by integrating out the heavy mode.
In Sec. \ref{sec:corr} we calculate the corrections to the power spectrum of the light mode by using the in-in formalism.
In Sec. \ref{sec:lar} We discuss the solutions of the coupled system on large scales.
In Sec. \ref{sec:con} we summarize our analysis.

\section{Perturbative analysis and the regularization} \label{sec:pert}
Let us consider the spatially covariant theories of gravity in  \cite{Gao:2019lpz}
\begin{equation}
S=\int\mathrm{d}t\mathrm{d}^{3}x\,N\sqrt{h}\left(a_{1}K+a_{2}F+b_{1}K_{ij}K^{ij}+b_{2}K^{2}+c_{1}KF+c_{2}F^{2}+d_{1}+d_{2}R\right)
,\label{action_1}\end{equation}
where
\begin{eqnarray}
	F:=\pounds_{\bm{n}}N=\frac{1}{N}\left(\dot{N}-\pounds_{\vec{N}}N\right), \label{F_def}
\end{eqnarray}
\begin{eqnarray}
K_{ij}:=\frac{1}{2}\pounds_{\bm{n}}h_{ij}=\frac{1}{2N}\left(\dot{h}_{ij}-\pounds_{\vec{N}}h_{ij}\right), \label{Kij_def}
\end{eqnarray}
the $K_{ij}$ is extrinsic curvature, and $N$ is lapse function.

The number of the degree of freedom of this model is 4: 2 scalar modes and 2 tensor modes. If we assume the lapse function to be an auxiliary variable, the theory should propagate 3 degrees of freedom. Generally there will be an extra scalar mode due to the presence of $F$ in the Lagrangian. In \cite{Gao:2019lpz}, the degenerate condition that leads to only one scalar field has been found. In the current work, we start from the general action (\ref{action_1}) and treat it as a model with two scalar modes, which is an analogue of two-field inflation models.

We express spacetime metric in term of the ADM variables

\begin{equation}
\mathrm{d}s^{2}=-N^{2}\mathrm{d}t^{2}+h_{ij}\left(\mathrm{d}x^{i}+N_{i}\mathrm{d}t\right)\left(\mathrm{d}x^{j}+N^{j}\mathrm{d}t\right),
\end{equation}
in this paper  we are concerned with  the scalar perturbation. We can separate the ADM variables to homogeneous part and
perturbations as
%Eq.(\ref{action_1})
\begin{equation}
N=e^{A},
\end{equation}
\begin{equation}
N_{i}=a\partial_{i}B,
\end{equation}
\begin{equation}
h_{ij}=a^{2}e^{2\zeta}\delta_{ij}.
\end{equation}
Eliminating the $B$ field that is non-dynamic  through the constraint equation, the  quadratic
action takes the form
\begin{align}
\label{action_scale}
S_{2}^{S} =& \int \mathrm{d}t\mathrm{d}^{3}x\,a^{3}\left(-w_{AA}A^{2}+g_{\zeta\zeta}\dot{\zeta}^{2}-\frac{1}{a^{2}}w_{\zeta\zeta}(\partial\zeta)^{2}+w_{A\zeta}A\dot{\zeta}\right.\\
 &\left.+g_{AA}\dot{A}^{2}+g_{A\zeta}\dot{A}\dot{\zeta}-\frac{1}{a^{2}}f_{A\zeta}\partial A\partial\zeta
 -\frac{1}{a^{2}}f_{AA}(\partial A)^{2}\right)\nonumber.
\end{align}

 The relation between the coefficients in (\ref{action_1}) and  coefficients in (\ref{action_scale}) is given in \cite{Gao:2019lpz}, and shown in \ref{app:coeff}.

Compared  to the  quadratic action of Horndeski theory, the $g_{AA}\dot{A}^{2}, g_{A\zeta}\dot{A}\dot{\zeta}$ are new-emerging terms. The degree of freedom of Horndeski theory is 3. But the degree of freedom in our model becomes 4 as the $A$ field becomes dynamic. The $\zeta$ field is a light field. When we treat the $g_{AA}\dot{A}^{2}, g_{A\zeta}\dot{A}\dot{\zeta}$ as perturbative terms, the kinetic  term is much smaller than the potential term, we think the $A$ field is a heavy field.
For the convenience of calculation, we make various coefficients $w_{AA}, g_{\zeta\zeta} , w_{\zeta\zeta}, w_{A\zeta}, g_{A\zeta}, g_{AA}, f_{A\zeta}, f_{AA}$ as constants that are not function of $t$.

It is now convenient to introduce the canonically normalized
variables

\begin{equation}
A=\alpha\frac{1}{a}\tilde{A},\zeta=\beta\frac{1}{a}\widetilde{\zeta},
\end{equation}
the coefficients satisfy
\begin{equation}
\alpha^{2}g_{AA}=\frac{1}{2},\beta^{2}g_{\zeta\zeta}=\frac{1}{2},
\end{equation}
and after using conformal time $\tau$ defined by $dt=ad\tau$, then
\begin{align}
\label{action_0}
S_{2}^{S} =&\int \mathrm{d}\tau\mathrm{d}^{3}x\,\left(-a^{2}\alpha^{2}w_{AA}\tilde{A}^{2}+\alpha^{2}\frac{a^{\prime\prime}}{a}g_{AA}\tilde{A}^{2}+\beta^{2}\frac{a^{\prime\prime}}{a}g_{\zeta\zeta}\tilde{\zeta}^{2}+\frac{1}{2}\tilde{\zeta}^{\prime}{}^{2}\right.\nonumber \\
 & +\frac{1}{2}\tilde{A}^{\prime}{}^{2}-\alpha^{2}f_{AA}\left(\partial\tilde{A}\right)^{2}-\beta^{2}w_{\zeta\zeta}\left(\partial\tilde{\zeta}\right)^{2}+\alpha\beta aw_{A\zeta}\tilde{A}\tilde{\zeta}^{\prime}-\alpha\beta f_{A\zeta}\left(\partial\tilde{A}\right)\left(\partial\tilde{\zeta}\right)\\
 &\left. -\alpha\beta a^{\prime}w_{A\zeta}\tilde{A}\tilde{\zeta}+\alpha\beta\frac{a^{\prime\prime}}{a}g_{A\zeta}\tilde{A}\tilde{\zeta}+\alpha\beta g_{A\zeta}\tilde{A}^{\prime}\tilde{\zeta}^{\prime}\right)\nonumber.
\end{align}

\section{Integrating the heavy mode} \label{sec:int}

The quadratic action for the scalar perturbations takes the form in (\ref{action_scale}).
In the case of GR or the case of usual spatially covariant gravity with one degree of freedom, the lapse function plays as an auxiliary variable with no dynamics. In other words, the coefficients $g_{AA}$ and $g_{A\zeta}$ are strictly vanishing. This can be equivalently viewed as that $A$ is infinitely heavy. For the purpose in this work, we relax this requirement by assuming a dynamical lapse function. Nevertheless, we assume the coefficients of the kinetic terms involving $A$ are small, i.e.,
\begin{equation}
\label{massH}
\left|\frac{g_{AA}}{w_{AA}}\right| \ll H^{-2}, \quad \left|\frac{g_{A\zeta}}{w_{AA}}\right| \ll H^{-2},
\end{equation}
the variation of $A$ is
\begin{equation}
\label{w_zeta}
-2w_{AA}A+w_{A\zeta}\partial_{t}\zeta-\left(6Hg_{AA}\dot{A}+2g_{AA}\ddot{A}+g_{A\zeta}\ddot{\zeta}-\frac{2}{a^{2}}f_{AA}\partial^{2}A-\frac{1}{a^{2}}f_{A\zeta}\partial^{2}\zeta\right)=0,
\end{equation}
 the mass term of the heavy field is satisfied
\begin{equation}
\label{mass}
m_{A}^{2}=\frac{w_{AA}}{g_{AA}},
\end{equation}
using (\ref{massH}), the $A$ can be treated as a heavy field with effective mass much larger than the Hubble scale.
Combining with (\ref{mass}), then the (\ref{w_zeta}) takes the form
\begin{equation}
-2m_{A}^{2}g_{AA}A+w_{A\zeta}\partial_{t}\zeta-\left(6Hg_{AA}\dot{A}+2g_{AA}\ddot{A}+g_{A\zeta}\ddot{\zeta}-\frac{2}{a^{2}}f_{AA}\partial^{2}A-\frac{1}{a^{2}}f_{A\zeta}\partial^{2}\zeta\right)=0,
\end{equation}
 for the $A$ field as a heavy mode, the terms $6Hg_{AA}\dot{A}+2g_{AA}\ddot{A}-\frac{2}{a^{2}}f_{AA}\partial^{2}A$ can be  neglected \cite{Garcia-Saenz:2019njm}, and we choose $f_{A\zeta}=0$ to simplify the calculation, which leads to a restriction on the coefficient $d_2$, shown in \ref{app:coeff} (for the case where the coefficient is not 0, we will discuss later). And using  the condition that $g_{A\zeta}$ is a small parameter, the relation between A and $\zeta$ results from solving its equation of motion at leading order as

\begin{equation}
A=\frac{w_{A\zeta}}{2w_{AA}}\partial_{t}\zeta,\label{A_zeta}
\end{equation}
substituting (\ref{A_zeta}) into the  action (\ref{action_scale}), and
neglecting the kinetic  gradient and kinetic terms of $A$, we obtain
the effective  Lagrangian
\begin{equation}
\mathcal{L}=\frac{w_{A\zeta}^{2}}{4w_{AA}}\left(\partial_{t}\zeta\right)^{2}+g_{\zeta\zeta}\left(\partial_{t}\zeta\right)^{2}-\frac{1}{a^{2}}w_{\zeta\zeta}\left(\partial\zeta\right)^{2}.
\end{equation}

 So after this treatment, the two fields becomes a
 effective single scalar field.
Compared to the action of the  free part of $\zeta$ field in (\ref{action_scale}),
 $\frac{w_{A\zeta}^{2}}{4w_{AA}}(\partial_{t}\zeta)^{2}$ is a new-emerging term. The action is
\begin{equation}
S_{2}^{S}=\int \mathrm{d}t\mathrm{d}^{3}x\,a^{3}\left(\frac{w_{A\zeta}^{2}}{4w_{AA}}\left(\partial_{t}\zeta\right)^{2}+g_{\zeta\zeta}\left(\partial_{t}\zeta\right)^{2}-\frac{1}{a^{2}}w_{\zeta\zeta}\left(\partial\zeta\right)^{2}\right),\label{action_zeta}
\end{equation}
it is now convenient to introduce the canonically normalized variables
\begin{equation}
\zeta=\gamma\frac{1}{a}\widetilde{\zeta},
\end{equation}
\hspace*{\fill} \\
\hspace*{\fill} \\
the coefficients satisfy
\begin{equation}
\gamma^{2}\left(g_{\zeta\zeta}+\frac{ w_{A\zeta}^{2}}{4w_{AA}}\right)=\frac{1}{2},
\end{equation}
then the action in (\ref{action_zeta}) will take the form
\begin{equation}
  S_{2}^{S}=\int \mathrm{d}\tau\mathrm{d}^{3}x\,\left(\frac{a^{\prime\prime}}{2a}\tilde{\zeta}^{2}+\frac{1}{2}\tilde{\zeta}^{\prime}{}^{2}-\gamma^{2}w_{\zeta\zeta}\left(\partial\zeta\right)^{2}\right),
 \end{equation}
the equation of motion for the mode function $u_{\zeta}$ takes the form

 \begin{equation}
\frac{\partial^{2}}{\partial\tau^{2}}u_{\zeta}\left(k,\tau\right)+2\left(k^{2}\gamma^{2}w_{\zeta\zeta}-\frac{a^{\prime\prime}}{2a}\right)u_{\zeta}\left(\tau,k\right)=0,\label{motion_zeta1}
\end{equation}
by seting $c_{\zeta}^{2}=2\gamma^{2}w_{\zeta\zeta}$, and solving in the de Sitter background, we find

\begin{align}
u_{\zeta}(\tau,k)=\frac{1}{\sqrt{2c_{\zeta}k}}\left(1-\frac{i}{c_{\zeta}k\tau}\right)e^{-ic_{\zeta}k\tau},
\end{align}
we calculate the power spectra in the usual way
\begin{equation}
\mathcal P_{\zeta}=\frac{k^{3}}{2\pi^{2}}\frac{\gamma^{2}}{a^{2}}\vert u_{\zeta}\vert^{2}.
\end{equation}

So the correction of the heavy field to the light field is equivalent to  a substitution of $\gamma$ to $\beta$, the solution to effective theory for the light mode $\zeta$  is still in the form of plane waves, but the  propagation speed of the $\zeta$ field has changed.

\section{Corrections of the heavy field to the light field} \label{sec:corr}
The $g_{AA}\dot{A}^{2}, g_{A\zeta}\dot{A}\dot{\zeta}$ are new-emerging perturbative terms. We use the  perturbation method of field theory to calculate the effect of the heavy $A$ field on the light $\zeta$ field.
For the convenience of calculation, we take $f_{A\zeta}=0,$ $w_{A\zeta}=0$ in action (\ref{action_0}), for this kind of non-small coupling terms are not easy to deal with. In our model, when the coefficients $g_{AA}$ and $g_{A\zeta}$ are not strictly vanishing, it is not easy to quantitatively evaluate the effect of the $f_{A\zeta}$ term. If the two fields are both light fields, it is not able  to solve motion equations analytically or to integrate out one field. For the two-field model with one light and one heavy modes (heavy enough), or one field is an auxiliary field, we find that the dispersion relation of $\zeta$ field will become complicated, such as appearance of  $k^4$ term. And in \cite{Fujita:2015ymn}, if $A$ is completely non-dynamic degree of freedom, and only if $f_{A\zeta}$ is constant, the contribution of the coefficient $f_{A\zeta}$ to the dispersion relation is mainly the ~$k^4$ term, so it mainly affects the mode with short wavelength, it will be only a correction to the overall magnitude of the power spectrum in large scales. After this treatment, the two fields are weakly coupled, the mode is still a two-field model that  a heavy field with a light field. When the two fields are  weakly coupled, the perturbative method of field theory can be used to deal with our model.
\begin{align}
\label{action_simplify}
S_{2}^{S} =&\int \mathrm{d}\tau\mathrm{d}^{3}x\,\left(-a^{2}\alpha^{2}w_{AA}\tilde{A}^{2}+\alpha^{2}\frac{a^{\prime\prime}}{a}g_{AA}\tilde{A}^{2}+\beta^{2}\frac{a^{\prime\prime}}{a}g_{\zeta\zeta}\tilde{\zeta}^{2}+\frac{1}{2}\tilde{\zeta}^{\prime}{}^{2}\right. \\
 &\left.+\frac{1}{2}\tilde{A}^{\prime}{}^{2}-\alpha^{2}f_{AA}\left(\partial\tilde{A}\right)^{2} -\beta^{2}w_{\zeta\zeta}\left(\partial\tilde{\zeta}\right)^{2}+\alpha\beta g_{A\zeta}\tilde{A}^{\prime}\tilde{\zeta}^{\prime}
 +\alpha\beta\frac{a^{\prime\prime}}{a}g_{A\zeta}\tilde{A}\tilde{\zeta}\right)\nonumber.
\end{align}

The perturbative terms are
\begin{equation}
\label{perturbative terms}
\alpha\beta g_{A\zeta}\tilde{A}^{\prime}\tilde{\zeta}^{\prime}+\alpha\beta\frac{a^{\prime\prime}}{a}g_{A\zeta}\tilde{A}\tilde{\zeta}.
\end{equation}

\subsection{Interaction Hamiltonian}

In the operator formalism of quantization, we should calculate Hamiltonian in the interaction picture. The Hamiltonian density which is defined by $H=\pi_{a}$$Q_{a}^{\prime}$-$\mathcal{L}$

\begin{equation}
\pi_{A}=\tilde{A}^{\prime}+g_{A\zeta}\alpha\beta\tilde{\zeta}^{\prime},\label{A_1}
\end{equation}

\begin{equation}
\pi_{\zeta}=\tilde{\zeta}^{\prime}+\alpha\beta g_{A\zeta}\tilde{A}^{\prime},\label{zeta_1}
\end{equation}
then the Hamiltonian can be written as

\begin{align}
H =&\frac{-\alpha^{2}\beta^{2}g_{A\zeta}w_{A\zeta}a\tilde{A}\pi_{A}}{\alpha^{2}\beta^{2}g_{A\zeta}^{2}-1}-\frac{\pi_{A}^{2}}{2\left(\alpha^{2}\beta^{2}g_{A\zeta}^{2}-1\right)}-\frac{\pi_{\zeta}^{2}}{2\left(\alpha^{2}\beta^{2}g_{A\zeta}^{2}-1\right)}\nonumber\\
 & +\frac{\alpha\beta g_{A\zeta}\pi_{A}\pi_{\zeta}}{\alpha^{2}\beta^{2}g_{A\zeta}^{2}-1}-\frac{\alpha^{2}\beta^{2}w_{A\zeta}^{2}a^{2}\tilde{A}^{2}}{2\left(\alpha^{2}\beta^{2}g_{A\zeta}^{2}-1\right)}+\frac{\alpha\beta w_{A\zeta}a\tilde{A}\pi_{\zeta}}{\alpha^{2}\beta^{2}g_{A\zeta}^{2}-1} \\
 & -\alpha\beta f_{A\zeta}k^{2}\tilde{A}\tilde{\zeta}+\alpha\beta a^{\prime}w_{A\zeta}\tilde{A}\tilde{\zeta}-\alpha\beta g_{A\zeta}\frac{a^{\prime\prime}}{a}\tilde{A}\tilde{\zeta}+\alpha^{2}f_{AA}k^{2}\tilde{A}^{2}\nonumber \\
 & +\alpha^{2}a^{2}w_{AA}\tilde{A}^{2}-\frac{a^{\prime\prime}}{2a}\tilde{A}^{2}-\frac{1}{2}\frac{a^{\prime\prime}}{a}\tilde{\zeta}^{2}+\beta^{2}w_{\zeta\zeta}k^{2}\tilde{\zeta}^{2},\nonumber \\
\nonumber
\end{align}
$H$ can be split into two parts, $H=H_{0}+H_{c}$, we take
the coupling coefficient $g_{A\zeta}$, $g_{AA}$as small parameters,
and$\ g_{A\zeta}$, $g_{AA}$ are the same order of magnitude, then

\begin{align}
H_{0}  =&\alpha^{2}f_{AA}k^{2}\tilde{A}^{2}+\alpha^{2}a^{2}w_{AA}\tilde{A}^{2}-\frac{a^{\prime\prime}}{2a}\tilde{A}^{2}+\frac{1}{2}\pi_{A}^{2}\\
 & -\frac{1}{2}\frac{a^{\prime\prime}}{a}\tilde{\zeta}^{2}+\beta^{2}w_{\zeta\zeta}k^{2}\tilde{\zeta}^{2}+\frac{1}{2}\pi_{\zeta}^{2}\nonumber, \end{align}
 \begin{equation}
  H_{c}=-\alpha\beta g_{A\zeta}\frac{a^{\prime\prime}}{a}\tilde{A}\tilde{\zeta}-\alpha\beta g_{A\zeta}\pi_{A}\pi_{\zeta}
\end{equation}
Where $H_{0}$ describes the free part while $H_{c}$
describes the cross interactions. From $H_{0}$, the free-theory canonical
momenta  are related with time-derivatives
of the fields as

\begin{equation}
\tilde{A}^{\prime}=\frac{\partial H_{0}}{\partial\pi_{A}}=\pi_{A},
\end{equation}

\begin{equation}
\tilde{\zeta}^{\prime}=\frac{\partial H_{0}}{\partial\pi_{\zeta}}=\pi_{\zeta},
\end{equation}
in interaction picture

\begin{align}
H_{0A}  =&\frac{1}{2}\tilde{A}^{\prime}{}^{2}+\alpha^{2}a^{2}w_{AA}\tilde{A}^{2}-\frac{a^{\prime\prime}}{2a}\tilde{A}^{2}\\
 & +\alpha^{2}f_{AA}k^{2}\tilde{A}^{2}\nonumber,
\end{align}

\begin{equation}
H_{0\zeta}=\frac{1}{2}\tilde{\zeta}^{\prime}{}^{2}-\frac{1}{2}\frac{a^{\prime\prime}}{a}\tilde{\zeta}^{2}+\beta^{2} w_{\zeta\zeta}k^{2}\tilde{\zeta}^{2},
\end{equation}

\begin{equation}
H_{c} =-\alpha\beta g_{A\zeta}\frac{a^{\prime\prime}}{a}\tilde{A}\tilde{\zeta}
  -\alpha\beta g_{A\zeta}\tilde{A}^{\prime}\tilde{\zeta}^{\prime},
\end{equation}
we can write $H_{c}$ as

\begin{equation}
H_{c}=f_{1}(\tau)\tilde{A}\tilde{\zeta}+f_{2}\tilde{A}^{\prime}\tilde{\zeta}^{\prime},\label{HC}
\end{equation}
$f_{1}$ is a function of $\tau$, $f_{2}$ is a constant.

To canonically quantize the system, the quantum fields are
decomposed as

\begin{equation}
\tilde A(\tau,k)=a_{k}u_{A}\left(\tau,k\right)+a_{-k}^{\dagger}u_{A}^{\ast}\left(\tau,k\right),
\end{equation}

\begin{align}
\tilde{\zeta}\left(k,\tau\right) & =b_{k}u_{\zeta}(k,\tau)+b_{-k}^{\dagger}u_{\zeta}^{\ast}\left(k,\tau\right),\\
\nonumber
\end{align}
the equations of motion for the mode functions $u_{A}$ and
$u_{\zeta}$ take the form

\begin{equation}
\frac{\partial^{2}}{\partial\tau^{2}}u_{A}\left(\tau,k\right)+2\left(\alpha^{2}a^{2}w_{AA}-\frac{a^{\prime\prime}}{2a}+\alpha^{2}f_{AA}k^{2}\right)u_{A}\left(\tau,k\right)=0,\label{motion_A}
\end{equation}

\begin{equation}
\frac{\partial^{2}}{\partial\tau^{2}}u_{\zeta}\left(k,\tau\right)+2\left(k^{2}\beta^{2}w_{\zeta\zeta}-\frac{a^{\prime\prime}}{2a}\right)u_{\zeta}\left(\tau,k\right)=0,\label{motion_zeta}
\end{equation}
then the mass of the heavy mode $m_{A}$ can be written as
\begin{equation}
m_{A}^{2}=2\alpha^{2}w_{AA},
\end{equation}
the equation (\ref{motion_A}) can be rewritten as
\begin{equation}
\frac{\partial^{2}}{\partial\tau^{2}}u_{A}\left(\tau,k\right)+\left(a^{2}m_{A}^{2}-\frac{a^{\prime\prime}}{a}+2\alpha^{2}f_{AA}k^{2}\right)u_{A}\left(\tau,k\right)=0.\label{motion_A1}
\end{equation}
Where $a$ and $\tau$ are related by $a=-\frac{1}{H\tau}$, it
is convenient to introduce new variable $x\equiv -\sqrt{2w_{\zeta\zeta}}\beta k\tau$ and $y\equiv-\sqrt{2f_{AA}}\alpha k\tau$, and we take  a special case that $w_{\zeta\zeta}=\frac{1}{2}$, $f_{AA}=\frac{1}{2}$. The solutions of (\ref{motion_zeta}), (\ref{motion_A1}) are
\begin{align}
\label{solution of zeta}
u_{\zeta}(x,k)=\frac{1}{\sqrt{2k\beta}}\left(1+\frac{i}{x}\right)e^{ix},
\end{align}
\begin{equation}
\label{solution of A}
u_{A}(y,k)=\frac{\sqrt{\pi}}{2}e^{-\frac{\pi}{2}\nu+i\frac{\pi}{4}}\sqrt{\frac{y}{k\alpha}}H_{i\nu}^{(1)}\left(y\right),
\end{equation}
\begin{equation}
\nu=\sqrt{\frac{m_{A}^{2}}{H^{2}}-\frac{9}{4}}.
\end{equation}

 We get Hamiltonian in the interaction picture, and for the part of free field, the solution to the heavy field is in the form of the Hankel function and the  solution to the light field is in the form of plane waves.

\subsection{ Using in-in formalism for the correction to spectra}
 We use the  field-theoretical perturbative method \cite{Gao:2009qy} to calculate the influence of the new field. For later convenience, we introduce $x = 1$ and $y = 1$ which are their respective values around sound horizon-crossings.

Now we would like to investigate the leading-order corrections to the power spectra. It is interesting to note that the leading-order correction is second-order correction  \cite{Gao:2009qy}.
Thus
\begin{equation}
\mathcal{P}_{x=1}=\mathcal{P}_{x=1}^{(0)}+\mathcal{P}_{x=1}^{(2)}.
\end{equation}
The two-point functions for $A$ and $\zeta$ are defined as
\begin{equation}
\left\langle \tilde{\zeta}\left(k_{1},\tau_{1}\right)\tilde{\zeta}\left(k_{2},\tau_{2}\right)\right\rangle ^{(0)}=\left(2\pi\right)^{3}\delta^{2}\left(k_{1}+k_{2}\right)\tilde{G_{k1}}\left(\tau_{1},\tau_{2}\right),
\end{equation}

\begin{equation}
\left\langle \tilde{A}\left(k_{1},\tau_{1}\right)\tilde{A}\left(k_{2},\tau_{2}\right)\right\rangle ^{(0)}=\left(2\pi\right)^{3}\delta^{2}\left(k_{1}+k_{2}\right)\tilde{F_{k1}}\left(\tau_{1},\tau_{2}\right),
\end{equation}
and

\begin{equation}
\tilde{G_{k1}}\left(\tau_{1},\tau_{2}\right)=u_{\zeta}\left(k,\tau\right)u_{\zeta}^{*}\left(k,\tau\right),
\end{equation}

\begin{equation}
\tilde{F_{k1}}\left(\tau_{1},\tau_{2}\right)=u_{A}\left(k,\tau\right)u_{A}^{*}\left(k,\tau\right).
\end{equation}
The relationship between the expected value of the field before and after the regular transformation
\begin{equation}
\left\langle {\zeta}\left(k_{1},\tau_{1}\right){\zeta}\left(k_{2},\tau_{2}\right)\right\rangle ^{(0)}=\left(2\pi\right)^{3}\delta^{2}\left(k_{1}+k_{2}\right){G_{k1}}\left(\tau_{1},\tau_{2}\right)
=\frac{{\beta}^{2}}{a^{2}}\left\langle\tilde{\zeta}\left(k_{1},\tau_{1}\right)\tilde{\zeta}\left(k_{2},\tau_{2}\right)\right\rangle ^{(0)}.
\end{equation}

When $g_{AA}$ is much smaller than $w_{AA}$, the model is  one heavy field with one light field.
 The power spectra for light and heavy fields are
\begin{equation}
\mathcal{P}_{A}^{(0)}=\frac{k^{3}\alpha^{2}}{2\pi^{2}}H^{2}\tau^{2}\vert u_{A}\vert^{2},
\end{equation}

\begin{equation}
\mathcal{P}_{\zeta}^{(0)}=\frac{k^{3}\beta^{2}}{2\pi^{2}}H^{2}\tau^{2}\vert u_{\zeta}\vert^{2},
\end{equation}
when $x=1$ at horizon-crossings, we obtain
\begin{equation}
\mathcal{P}_{\zeta}^{(0)}=\frac{1}{2\beta\pi^{2}}H^{2}.
\end{equation}

We only calculate the correction of the heavy field on the light field. The second-order correction is the leading order and in the form
\begin{align}\label{O2}\left\langle \hat{O}\left(n\right)\right\rangle ^{(2)}= & -2\Re\int_{-\infty^{+}}^{\tau}d\tau^{\prime}\int_{-\infty^{+}}^{\tau^{\prime}}d\tau^{\prime\prime}\left\langle 0_{I}|\hat{O_{I}}\left(n\right)H_{c}(\tau^{\prime})H_{c}\left(\tau^{\prime\prime}\right)|0_{I}\left(n\right)\right\rangle \\
 & +\int_{-\infty^{-}}^{\tau}d\tau^{\prime}\int_{-\infty^{+}}^{\tau}d\tau^{\prime\prime}\left\langle 0_{I}|H_{c}(\tau^{\prime})\hat{O_{I}}\left(n\right)H_{c}\left(\tau^{\prime\prime}\right)|0_{I}\left(n\right)\right\rangle \nonumber,
\end{align}
according to (\ref{HC}), (\ref{O2})
, we calculate the second-order correction of the $A$ field to the $\zeta$ field
\begin{align}
\label{correction of zeta to A}\left\langle {\zeta}\left(\tau,k_{1}\right){\zeta}\left(\tau,k_{2}\right)\right\rangle ^{(2)}= & \left(2\pi\right)^{3}\delta^{3}(k_{1}+k_{2})\times\Big\{-4\Re\int_{-\infty}^{\tau}d\tau^{\prime}\int_{-\infty}^{\tau^{\prime}}d\tau^{\prime\prime}\nonumber\\
 & \Big[f_{2}^{2}\frac{d}{d\tau^{\prime}}G\left(\tau,\tau^{\prime}\right)\frac{d}{d\tau^{\prime\prime}}G\left(\tau,\tau^{\prime\prime}\right)\frac{d}{d\tau^{\prime}}\frac{d}{d\tau^{\prime\prime}}F\left(\tau^{\prime},\tau^{\prime\prime}\right)+f_{2}f_{1}\left(\tau^{\prime\prime}\right)\frac{d}{d\tau^{\prime}}G\left(\tau,\tau^{\prime}\right)G\left(\tau,\tau^{\prime\prime}\right)\frac{d}{d\tau^{\prime}}F\left(\tau^{\prime},\tau^{\prime\prime}\right)\nonumber\\
 & +f_{1}\left(\tau^{\prime}\right)f_{2}G\left(\tau,\tau^{\prime}\right)\frac{d}{d\tau^{\prime\prime}}G\left(\tau,\tau^{\prime\prime}\right)\frac{d}{d\tau^{\prime\prime}}F\left(\tau^{\prime},\tau^{\prime\prime}\right)+f_{1}(\tau^{\prime})f_{1}\left(\tau^{\prime\prime}\right)G\left(\tau,\tau^{\prime}\right)G\left(\tau,\tau^{\prime\prime}\right)F\left(\tau^{\prime},\tau^{\prime\prime}\right)\Big]\nonumber\\
 & +2\int_{-\infty}^{\tau}d\tau^{\prime}\int_{-\infty}^{\tau}d\tau^{\prime\prime}\Big[f_{2}^{2}\frac{d}{d\tau^{\prime}}G\left(\tau^{\prime},\tau\right)\frac{d}{d\tau^{\prime\prime}}G\left(\tau,\tau^{\prime\prime}\right)\frac{d}{d\tau^{\prime}}\frac{d}{d\tau^{\prime\prime}}F\left(\tau^{\prime},\tau^{\prime\prime}\right)\\
 & +f_{2}f_{1}\left(\tau^{\prime\prime}\right)\frac{d}{d\tau^{\prime}}G\left(\tau^{\prime},\tau\right)G\left(\tau,\tau^{\prime\prime}\right)\frac{d}{d\tau^{\prime}}F\left(\tau^{\prime},\tau^{\prime\prime}\right)+f_{1}\left(\tau^{\prime}\right)f_{2}\frac{d}{d\tau^{\prime\prime}}G\left(\tau,\tau^{\prime\prime}\right)G\left(\tau^{\prime},\tau\right)\frac{d}{d\tau^{\prime\prime}}F\left(\tau^{\prime},\tau^{\prime\prime}\right)\nonumber\\
 & +f_{1}\left(\tau^{\prime}\right)f_{1}\left(\tau^{\prime\prime}\right)G\left(\tau,\tau^{\prime\prime}\right)G\left(\tau^{\prime},\tau)F(\tau^{\prime},\tau^{\prime\prime}\right)\Big]\Big\}\nonumber.
\end{align}
The leading-order correction to the power spectra of $\zeta$ can be denoted as
\begin{align}
 \left\langle {\zeta}\left(\tau,k_{1}\right){\zeta}\left(\tau,k_{2}\right)\right\rangle^{(2)}=&
 \left(2\pi\right)^{3}\delta^{3}(k_{1}+k_{2})\mathcal{P}_{\zeta}^{(2)},
\end{align}
that
\begin{equation}
\label{r}
r=\frac{\alpha}{\beta}=\sqrt{\frac{g_{\zeta\zeta}}{g_{AA}}},
\end{equation}
inserting (\ref{solution of zeta}), (\ref{solution of A}) into (\ref{correction of zeta to A}), and use the relationship between $x$ and $\tau$, the second-order correction is
\begin{equation}
\mathcal{P}_{\zeta}^{(2)}=r^{2}e^{-\pi\nu}\frac{H^{2} g_{A\zeta}^{2}}{8k^{3}{\beta}}\Gamma\left(x,r\right),\label{P_2}
\end{equation}
that
\begin{align}\label{double integral}
\Gamma(x,r)= & -4\Re\left(1+\frac{i}{x}\right)e^{ix}\left(1+\frac{i}{x}\right)e^{ix}\times\int_{x}^{\infty}dx^{\prime}\int_{x^{\prime}}^{\infty}dx^{\prime\prime}\nonumber\\
 & \left(\frac{\partial}{\partial x^{\prime}}\left(x^{\prime^{\frac{3}{2}}}H_{i\nu}^{(1)}\left(x^{\prime}r\right)\right)\frac{\partial}{\partial x^{\prime}}\left(\left(x^{\prime}-i\right)e^{-ix^{\prime}}\right)+2\sqrt{x^{\prime}}H_{i\nu}^{(1)}\left(x^{\prime}r\right)\left(1-\frac{i}{x^{\prime}}\right)e^{-ix^{\prime}}\right)\nonumber\\
 & \times\left(\frac{\partial}{\partial x^{\prime\prime}}\left(x^{\prime\prime^{\frac{3}{2}}}H_{i\nu}^{(1)\ast}\left(x^{\prime\prime}r\right)\right)\frac{\partial}{\partial x^{\prime\prime}}\left(\left(x^{\prime\prime}-i\right)e^{-ix^{\prime\prime}}\right)+2\sqrt{x^{\prime\prime}}H_{i\nu}^{(1)\ast}\left(x^{\prime\prime}r\right)\left(1-\frac{i}{x^{\prime\prime}}\right)e^{-ix^{\prime\prime}}\right)\\
 & +2\left(1-\frac{i}{x}\right)e^{-ix}\left(1+\frac{i}{x}\right)e^{ix}\times\int_{x}^{\infty}dx^{\prime}\int_{x}^{\infty}dx^{\prime\prime}\nonumber\\
 & \left(\frac{\partial}{\partial x^{\prime}}\left(x^{\prime^{\frac{3}{2}}}H_{i\nu}^{(1)}\left(x^{\prime}r\right)\right)\frac{\partial}{\partial x^{\prime}}\left(\left(x^{\prime}+i\right)e^{ix^{\prime}}\right)+2\sqrt{x^{\prime}}H_{i\nu}^{(1)}\left(x^{\prime}r\right)\left(1+\frac{i}{x^{\prime}}\right)e^{ix^{\prime}}\right)\nonumber\\
 & \times\left(\frac{\partial}{\partial x^{\prime\prime}}\left(x^{\prime\prime^{\frac{3}{2}}}H_{i\nu}^{(1)\ast}\left(x^{\prime\prime}r\right)\right)\frac{\partial}{\partial x^{\prime\prime}}\left(\left(x^{\prime\prime}-i\right)e^{-ix^{\prime\prime}}\right)+2\sqrt{x^{\prime\prime}}H_{i\nu}^{(1)\ast}\left(x^{\prime\prime}r\right)\left(1-\frac{i}{x^{\prime\prime}}\right)e^{-ix^{\prime\prime}}\right)\nonumber,
\end{align}
and we define
\begin{equation}
\Gamma_{r}=r^{2}e^{-\pi\nu}\Gamma(1,r),\label{Gamma_r}
\end{equation}
the equation of (\ref{double integral}) is a double integral, we will calculate its value at $x=1$. We find  variables except $r$ can be extracted outside the double integral, so calculating $\Gamma_{r}$ becomes the key to calculating the second-order correction. It is difficult to get an analytical solution, we can only do numerical integration.

Due to the $g_{A\zeta}$ is small coefficient, the second-order correction in (\ref{P_2}) is a relatively small value. We find that the magnitude of this second-order correction is related to the parameters  $\beta, \nu, k,r$. The $r$ value reflects the relationship between the kinetic energy coefficients of the $A$ field and the $\zeta$ field, the value of $r$ in (\ref{r}) should be quite greater than 1. And the perturbative terms in (\ref{perturbative terms})  are related to ratio $r$.
\begin{figure}[H]
\centering
\includegraphics[width=20pc]{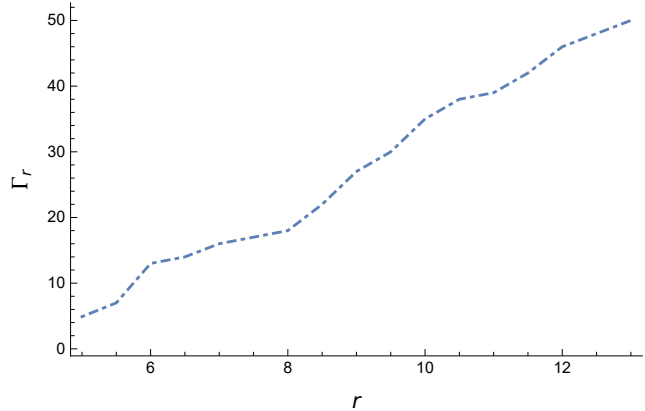}
\caption{The  $\Gamma_{r}$ is defined in (\ref{Gamma_r}).  We take points within the range of r equal to 5 to 13.} \label{plot60}
\end{figure}
We find that this correction  is highly correlated  with the ratio $r$, the $r$ value reflects the relationship between the kinetic energy coefficients of the $A$ field and the $\zeta$ field. We conclude that the $\Gamma_{r}$  increase with the ratio $r$.

%\begin{figure}[H]
%\centering
%\includegraphics[width=19pc]{plot1.eps}
%\caption{The  $\Gamma_{r}$ is defined in (\ref{Gamma_r}).  We take points within %the range of r equal to 5 to 25 and  connect them with polylines.}
 %\label{plot1}
%\end{figure}

\section{Solution in large scales} \label{sec:lar}

When using the perturbative method of field theory, the coefficients of the coupling terms in action (\ref{action_scale})  are small  or take the value of 0. In this section, we  deal with the case that the coupling coefficient $w_{A\zeta}$  is not a small value.
 We find  analytical solutions to the coupled system in large scale. In \cite{Tolley:2007nq,Koyama:2007ag}, calculating the spectrum  under the limit $k \ensuremath{\ll} aH$ has been presented.

From the action (\ref{action_0}), the equations of motion for the mode functions $u_{A}$ and $u_{\zeta}$  take the form
\begin{align}
&
-2a^{2}\alpha^{2}w_{AA}u_{A}+2\alpha^{2}\frac{a^{\prime\prime}}{a}g_{AA}u_{A}-\alpha\beta a^{\prime}w_{A\zeta}u_{\zeta}+\alpha\beta aw_{A\zeta}u_{\zeta}^{\prime}\nonumber\\&=\frac{\partial^{2}}{\partial\tau^{2}}u_{A}+k^{2}\left(\alpha\beta f_{A\zeta}u_{\zeta}+2\alpha^{2}f_{AA}u_{A}\right)+\alpha\beta g_{A\zeta}u_{\zeta}^{\prime\prime},\label{motion_1}
\end{align}

\begin{align}
&
\frac{a^{\prime\prime}}{a}u_{\zeta}-\alpha\beta a^{\prime}w_{A\zeta}u_{A}+\alpha\beta\frac{a^{\prime\prime}}{a}g_{A\zeta}u_{A}-\alpha\beta a^{\prime}w_{A\zeta}u_{A}-\alpha\beta aw_{A\zeta}u_{A}^{\prime}\nonumber\\&=\frac{\partial^{2}}{\partial\tau^{2}}u_{\zeta}+k^{2}\left(2\beta^{2}w_{\zeta\zeta}u_{\zeta}+\alpha\beta f_{A\zeta}u_{A}\right)+\alpha\beta g_{A\zeta}u_{A}^{\prime\prime}.\label{motion_2}
\end{align}

When $a=-\frac{1}{H\tau}$, the condition $k\ll aH$  leads to $k\tau\rightarrow0$, assuming that the equation has a power law solution , meanwhile $u_{A}$ and $u_{\zeta}$ are same dependence on $\tau$.  We assume that the solution to the equation in the following form \cite{Tolley:2007nq}
\begin{equation}
u_{A}=B\tau^{p},\label{sol_1}
\end{equation}

\begin{equation}
u_{\zeta}=C\tau^{p},\label{sol_2}
\end{equation}
then the terms include $k^{2}$ terms can be ignored, substituting ({\ref{sol_1}), ({\ref{sol_2}) into  ({\ref{motion_1}), ({\ref{motion_2}), we obtain

\begin{equation}
-2\frac{1}{H^{2}}\alpha^{2}w_{AA}B+2B-\alpha\beta\frac{1}{H}w_{A\zeta}C-\alpha\beta\frac{1}{H}w_{A\zeta}pC=p\left(p-1\right)B+\alpha\beta g_{A\zeta}p\left(p-1\right)C
,\label{sol_3}\end{equation}

\begin{equation}
2C-2\alpha\beta\frac{1}{H}w_{A\zeta}B+2\alpha\beta g_{A\zeta}B=p\left(p-1\right)C-\alpha\beta\frac{1}{H}w_{A\zeta}pB+\alpha\beta g_{A\zeta}p\left(p-1\right)B
.\label{sol_4}\end{equation}
From  (\ref{sol_3}), (\ref{sol_4}), we can get the relationship
between $B$ and $C$ and value of $p$. Then we use $g_{A\zeta}$ as a small parameter to simplify calculation, there are two sets of solutions, one is
\begin{equation}
p_{1}=2,
\end{equation}
\begin{equation}
B_{1}/C_{1}=-\frac{3\beta Hw_{A\zeta}}{2\alpha w_{AA}},
\end{equation}
the other is
\begin{equation}
\label{P_value}
p_{2}=1/2+\left(9-8\frac{1}{H^{2}}\alpha^{2}w_{AA}-4\alpha^{2}\beta^{2}\frac{1}{H^{2}}w_{A\zeta}^{2}\right)^{1/2},
\end{equation}
\begin{equation}
B_{2}/C_{2}=\frac{p_{2}+1}{\alpha\beta\frac{1}{H}w_{A\zeta}},
\end{equation}
\begin{equation}
\label{P_value1}
p_{3}=1/2-\left(9-8\frac{1}{H^{2}}\alpha^{2}w_{AA}-4\alpha^{2}\beta^{2}\frac{1}{H^{2}}w_{A\zeta}^{2}\right)^{1/2},
\end{equation}
\begin{equation}
B_{3}/C_{3}=\frac{p_{3}+1}{\alpha\beta\frac{1}{H}w_{A\zeta}}.
\end{equation}

 We find the value of $p$ is constant or only depends on some coefficients of the model and $H$, and the $p$ has 3 solutions and the corresponding $B/C$ also has 3 solutions.
Then

\begin{equation}
u_{A}=B_{1}k^{p_{1}-3/2}\left(-\tau\right)^{p_{1}}+B_{2}k^{p_{2}-3/2}\left(-\tau\right)^{p_{2}}+B_{3}k^{p_{3}-3/2}\left(-\tau\right)^{p_{3}},
\end{equation}
\begin{equation}
u_{\zeta}=C_{1}k^{p_{1}-3/2}\left(-\tau\right)^{p_{1}}+C_{2}k^{p_{2}-3/2}\left(-\tau\right)^{p_{2}}+C_{3}k^{p_{3}-3/2}\left(-\tau\right)^{p_{3}},
\end{equation}
we calculate the power spectra in the usual way

\begin{equation}
\mathcal P_{\zeta}=\frac{k^{3}}{2\pi^{2}}\frac{\beta^{2}}{a^{2}}\vert u_{\zeta}\vert^{2},
\end{equation}
for de Sitter background, the spectral index is

 \begin{equation}
n_{s}=2p+1.
\end{equation}
We find that for the solutions of $p1, p2, p3$, the spectral indices obtained by $p1$ and $p2$ are obviously greater than 1, only $p3$ can achieve a spectral index close to 1, then the coefficients should satisfy
\begin{equation}
8\frac{1}{H^{2}}\alpha^{2}w_{AA}+4\alpha^{2}\beta^{2}\frac{1}{H^{2}}w_{A\zeta}^{2}\approx\frac{35}{4}.
\end{equation}

From (\ref{P_value1}), we can change the value of $p$  by adjusting the parameters in the model, a good $p$ solution can make spectral index  close to 1 in large scales.

\section{Conclusions} \label{sec:con}
The spatially covariant gravity with a dynamical lapse function can naturally be viewed as a model with two scalar modes, which is an analogue of two-field inflation model.

In this paper, by regarding the  newly-emerging terms  as the perturbative terms, we  integrate out the heavy mode $A$, which arises due to the dynamical lapse function,  and get the effective theory for the light mode $\zeta$, which is the usual curvature perturbation. We find that the propagation speed of $\zeta$ field is modified due to the presence of $A$. When the two fields are  weakly coupled, the perturbative method of field theory can be used to deal with our model. For the part of free field, we find that the solution to the heavy mode is in the form of Hankel function, and for the light mode the solution is in the form of plane waves.
Then we employ the standard in-in formalism to calculate the leading-order corrections to the power spectra of $\zeta$ from the heavy mode $A$.
 Finally, when the two fields are not weakly coupled  that the $w_{A\zeta}$ is not a small coffecient, we obtain  a power law solution as the form  $\tau^{p}$ in large scales, the index of $p$ has three  solutions, we find that only $p3$ can lead to a spectral index close to 1.

\section*{Acknowledgements}

This work was partly supported by the Natural Science Foundation of China (NSFC) under the grant No. 11975020.

\appendix

\section{Coefficients} \label{app:coeff}
For the action in (\ref{action_1}), the quadratic action for the scalar modes takes the general form
\begin{equation}
	S_{2}^{\mathrm{S}}\left[\zeta,A,B\right]=\int\mathrm{d}t\mathrm{d}^{3}x\,a^{3}\left(L_{2}^{(1)}+L_{2}^{(2)}\right),\label{S2_S_ori}
	\end{equation}

\begin{eqnarray}
	L_{2}^{(1)} & = & \mathcal{C}_{\dot{\zeta}^{2}}\dot{\zeta}^{2}+\mathcal{C}_{\dot{\zeta}\dot{A}}\dot{\zeta}\dot{A}+\mathcal{C}_{\dot{A}^{2}}\dot{A}^{2}\nonumber \\
	&  & -\mathcal{C}_{\dot{\zeta}B}\dot{\zeta}\frac{\partial^{2}B}{a} -\mathcal{C}_{\dot{A}B}\dot{A}\frac{\partial^{2}B}{a}+\mathcal{C}_{B^{2}}\frac{(\partial^{2}B)^{2}}{a^{2}},
	\end{eqnarray}
	
terms which are irrelevant to counting the number of degrees of freedom are
	\begin{eqnarray}
	L_{2}^{(2)} & = & \mathcal{C}_{\zeta^{2}}\zeta^{2}+\mathcal{C}_{\dot{\zeta}A}\dot{\zeta}A+\mathcal{C}_{\zeta A}\zeta A+\mathcal{C}_{A^{2}}A^{2} - \mathcal{C}_{AB}A\frac{\partial^{2} B}{a}, \label{L2^2}
	\end{eqnarray}
the following coefficients are by Taylor expansion expressed with coefficients in (\ref{action_1}), for example, $a_{i}$ is actually $a_{i}\left(t,N\right)|_{N=1}$, $b_{i}$, $c_{i}$, $d_{i}$ are similar cases, the coefficients are
	\begin{eqnarray}
	\mathcal{C}_{\dot{\zeta}^{2}} & = & 3(b_{1}+3b_{2}),\\
	\mathcal{C}_{\dot{\zeta}\dot{A}} &= & 3c_{1},\\
	\mathcal{C}_{\dot{A}^{2}} & = & c_{2},\\
	\mathcal{C}_{\dot{\zeta}B} & = & 2(b_{1}+3b_{2}),\\
	\mathcal{C}_{\dot{A}B} & = & c_{1},\\
	\mathcal{C}_{B^{2}} & = & b_{1}+b_{2},
	\end{eqnarray}

\begin{equation}
	\mathcal{C}_{\zeta^{2}}=-2d_{2}\frac{\partial^{2}}{a^{2}},
	\end{equation}
	\begin{equation}
	\mathcal{C}_{\dot{\zeta}A}=-3H\left[2\left(b_{1}+3b_{2}\right)-2\frac{\partial\left(b_{1}+3b_{2}\right)}{\partial N}+3c_{1}\right]+3\left(\frac{\partial a_{1}}{\partial N}-a_{2}\right),
	\end{equation}
	\begin{equation}
	\mathcal{C}_{\zeta A} = -4\left(d_{2}+\frac{\partial d_{2}}{\partial N}\right)\frac{\partial^{2}}{a^{2}},
	\end{equation}
	\begin{eqnarray}
	\mathcal{C}_{A^{2}} & = & -\frac{3}{2}\frac{\partial c_{1}}{\partial N}\dot{H}+\frac{1}{2}\left(d_{1}+3\frac{\partial d_{1}}{\partial N}+\frac{\partial^{2}d_{1}}{\partial N^{2}}\right)\nonumber \\
	&  & +\frac{3}{2}H^{2}\left[b_{1}+3b_{2}-\frac{\partial\left(b_{1}+3b_{2}\right)}{\partial N}+\frac{\partial^{2}\left(b_{1}+3b_{2}\right)}{\partial N^{2}}-3\frac{\partial c_{1}}{\partial N}\right]\nonumber \\
	&  & +\frac{3}{2}H\left(\frac{\partial a_{1}}{\partial N}-a_{2}+\frac{\partial^{2}a_{1}}{\partial N^{2}}-\frac{\partial a_{2}}{\partial N}\right)\nonumber \\
	&  & -\frac{1}{2}\left[3H\frac{\partial a_{1}}{\partial X}+3\frac{\partial\left(b_{1}+3b_{2}\right)}{\partial X}H^{2}+\frac{\partial d_{1}}{\partial X}\right]\frac{\partial^{2}}{a^{2}},
	\end{eqnarray}
	\begin{equation}
	\mathcal{C}_{AB}=\frac{1}{3}\mathcal{C}_{\dot{\zeta}A}.
	\end{equation}

Eliminating the $B$ field that is non-dynamic through the constraint equation, the quadratic action takes the form

\begin{align}
S_{2}^{S} =& \int \mathrm{d}t\mathrm{d}^{3}x\,a^{3}\left(-w_{AA}A^{2}+g_{\zeta\zeta}\dot{\zeta}^{2}-\frac{1}{a^{2}}w_{\zeta\zeta}(\partial\zeta)^{2}+w_{A\zeta}A\dot{\zeta}\right.\\
 &\left.+g_{AA}\dot{A}^{2}+g_{A\zeta}\dot{A}\dot{\zeta}-\frac{1}{a^{2}}f_{A\zeta}\partial A\partial\zeta
 -\frac{1}{a^{2}}f_{AA}(\partial A)^{2}\right)\nonumber.
\end{align}

The coefficients for the kinetic terms can be evaluated explicitly to be
	\begin{eqnarray}
	g_{\zeta\zeta} & = & \frac{2b_{1}(b_{1}+3b_{2})}{b_{1}+b_{2}},\\
	g_{A\zeta} & = & \frac{2b_{1}c_{1}}{b_{1}+b_{2}},\\
	g_{AA} & = & c_{2}-\frac{c_{1}^{2}}{4(b_{1}+b_{2})},
	\end{eqnarray}

\begin{equation}
	w_{\zeta\zeta}=-2d_{2},
	\end{equation}

\begin{equation}
w_{A\zeta}=  \mathcal{C}_{\dot{\zeta}A}-\frac{1}{2}\frac{\mathcal{C}_{\dot{\zeta}B}\mathcal{C}_{AB}}{\mathcal{C}_{B^{2}}},
\end{equation}

\begin{equation}
f_{AA} = -\frac{1}{2}\left[3H\frac{\partial a_{1}}{\partial X}+3\frac{\partial\left(b_{1}+3b_{2}\right)}{\partial X}H^{2}+\frac{\partial d_{1}}{\partial X}\right],
\end{equation}

\begin{align}
	w_{AA} = & \frac{3}{2}\frac{\partial c_{1}}{\partial N}\dot{H}-\frac{1}{2}\left(d_{1}+3\frac{\partial d_{1}}{\partial N}+\frac{\partial^{2}d_{1}}{\partial N^{2}}\right)\nonumber \\
	&   -\frac{3}{2}H^{2}\left[b_{1}+3b_{2}-\frac{\partial\left(b_{1}+3b_{2}\right)}{\partial N}+\frac{\partial^{2}\left(b_{1}+3b_{2}\right)}{\partial N^{2}}-3\frac{\partial c_{1}}{\partial N}\right]\nonumber \\
	&   -\frac{3}{2}H\left(\frac{\partial a_{1}}{\partial N}-a_{2}+\frac{\partial^{2}a_{1}}{\partial N^{2}}-\frac{\partial a_{2}}{\partial N}\right)\nonumber \\
& 	+\frac{1}{4}\frac{\mathcal{C}_{AB}^{2}}{\mathcal{C}_{B^{2}}}-\frac{1}{4}\frac{1}{a^{3}}\partial_{t}\left(a^{3}\frac{\mathcal{C}_{AB}\mathcal{C}_{\dot{A}B}}{\mathcal{C}_{B^{2}}}\right),
\end{align}
\begin{equation}
f_{A\zeta} = -4\left(d_{2}+\frac{\partial d_{2}}{\partial N}\right),
\end{equation}
for the coefficient $f_{A\zeta}$ is equal to 0,  we get a relation that $-4\left(d_{2}+\frac{\partial d_{2}}{\partial N}\right)|_{N=1}=0$. There is no general solution to this case, one of the solutions is $d_2(t,N)=c-c\frac{N}{2}$, where $c$ is a constant.

\bibliographystyle{aapmrev4-2}
\bibliography{Bibfile}

%\bibliography{C:/Dropbox/_WORK/_GENERIC/BIBTEX/Articles,C:/Dropbox/_WORK/_GENERIC/BIBTEX/Books}

\end{document}